\documentclass[twocolumn]{aastex631}

\graphicspath{{./}{figures/}}

\begin{document}

\title{New Close Binary Central Stars of Planetary Nebulae from \textit{Gaia} DR3 Epoch Photometry}

\author[0000-0002-8767-3907]{Nicholas Chornay}
\affiliation{Institute of Astronomy, University of Cambridge, Madingley Road, Cambridge CB3 0HA, UK}
\correspondingauthor{Nicholas Chornay}
\email{njc89@cam.ac.uk}

\author[0000-0003-3983-8778]{Nicholas A. Walton}
\affiliation{Institute of Astronomy, University of Cambridge, Madingley Road, Cambridge CB3 0HA, UK}

\begin{abstract}

Close binary interactions perform a key role in the formation and shaping of planetary nebulae. However only a small fraction of Galactic planetary nebulae are known to host close binary systems. Many such systems are detectable through photometric variability. We searched recently published epoch photometry data from \textit{Gaia} DR3 for planetary nebula central stars with periodic photometric variability indicative of binarity, uncovering four previously unknown close binaries.

\end{abstract}


\section{Introduction}

A significant fraction of planetary nebulae (PNe) have central stars (CSPNe) that evolved in close binary systems \citep{jonesboffin2017}. However only around a hundred such systems are known out of a few thousand PNe \citep[though estimates of photometrically detectable fraction are on the order of 20\%:][]{miszalski2009oglebinaries,jacoby2021kepler}. Expanding the sample further will improve statistical significance and allow stronger conclusions to be drawn about the properties of these objects.

Many close binary CSPNe have been identified from their photometric variability \citep[due to irradiation, ellipsoidal modulation, and eclipses;][]{boffinjones2019book}, both as a result of dedicated monitoring \citep[e.g.][]{corradi2011necklace} and in data from larger surveys \citep[e.g.][]{miszalski2009oglebinaries}. More recently new binaries have been found using space-based photometry from missions such as Kepler/K2 \citep{jacoby2021kepler} and \textit{Gaia} \citep{chornay2021gaiabinaries}. In this work we report on new discoveries from the epoch (time series) photometry released as part of \textit{Gaia} Data Release 3 \citep[DR3;][]{gaia2022dr3arxiv}.

\section{Methods}

Epoch photometric data is published in \textit{Gaia} DR3 for objects classified as variable in its data processing \citep{eyer2022variablesarxiv}. We cross-matched the CSPN catalog of \citet{chornay2021cspn} (including unpublished low-confidence matches for completeness) with the variable sources in \textit{Gaia} DR3 and retrieved epoch photometry for the 126 sources that resulted from the cross-match. As was done in \citet{chornay2021gaiabinaries}, we performed a Lomb-Scargle analysis \citep[using the implementation of][]{vanderplas2015} on the $G$-band light curves in order to search for periodic variability on timescales between hours and weeks. The resulting periodograms and folded light curves were inspected for indications of close binarity.

\section{Results}

We identified four CSPNe whose \textit{Gaia} DR3 light curves show strong signatures of close binarity, and which were not previously known to be binaries.\footnote{Based on the list compiled at \url{https://www.drdjones.net/bcspn/}, as of 24th August 2021.} There are other CSPNe that exhibit periodic variability in the \textit{Gaia} data, but for which the origin is less clear, due to high scatter (e.g.~Kn 51), low amplitude (e.g.~NGC 6891), unclear association of the source with the PN (e.g.~BRAN 229), or sparse sampling. These are left for future work.

\begin{figure*}
\includegraphics[width=\textwidth]{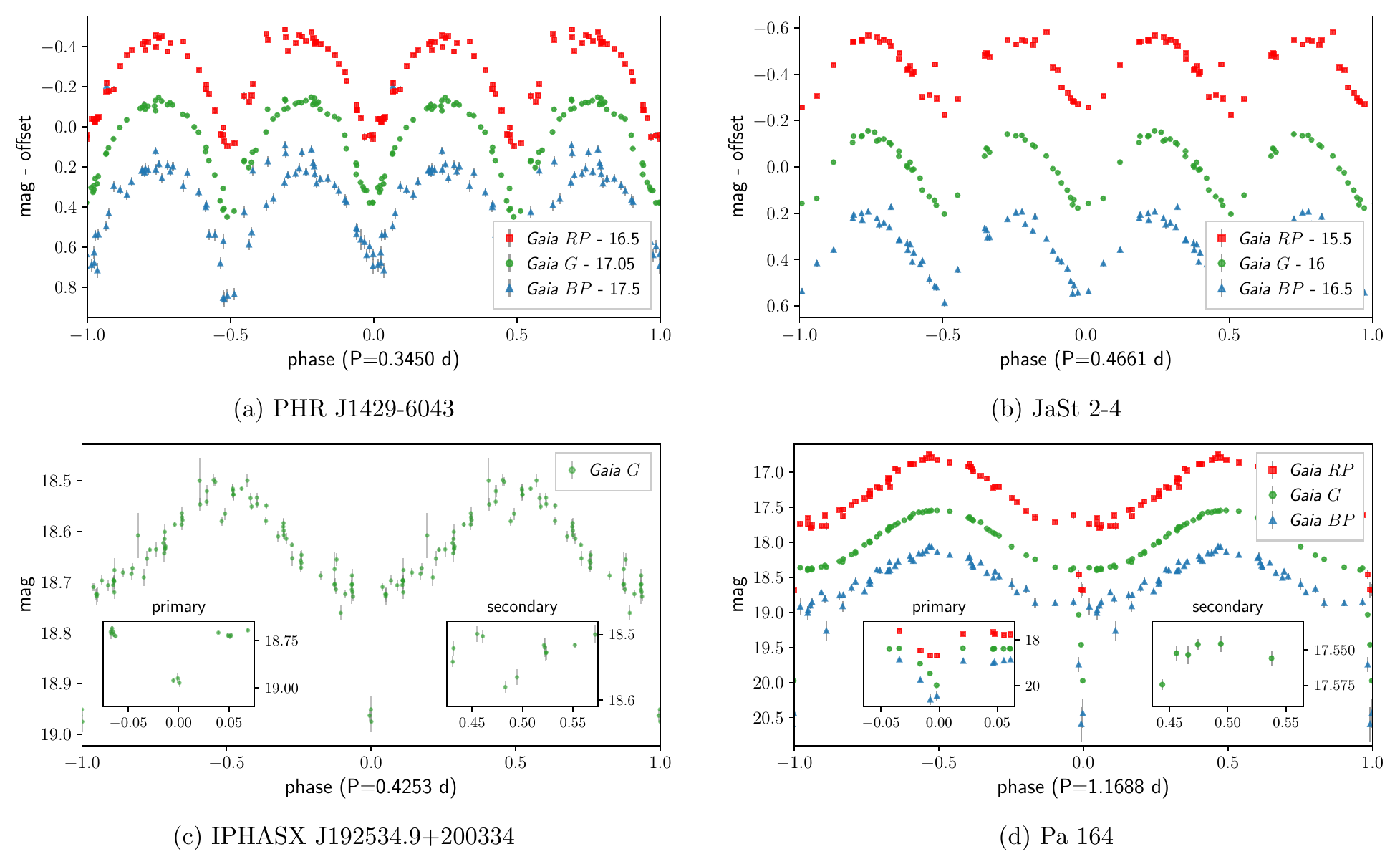}
\caption{Folded \textit{Gaia} light curves for the four close binary systems identified in this work.
\label{fig:binaries}}
\end{figure*}

\textit{PHR J1429--6043:} this is a faint, extended ``possible" PN (PN G314.6--00.1) from the Macquarie/AAO/Strasbourg H$\alpha$ PNe catalogue \citep[MASH;][]{mashpn}. The identity of the CSPN in \textit{Gaia} has been uncertain, with \citet{chornay2021cspn} not publishing a match. \textit{Gaia} DR3 reveals that the closest source to the center of the nebula
(\textit{Gaia} DR3 5878618052702285056, $\alpha = 14^h29^m53.08^s\ \delta = -60^\circ43^{\prime\prime}56.25^\prime$)
is indeed the CSPN,
an ellipsoidally modulated system with a $\sim$8.3 hour orbital period (Fig.~\ref{fig:binaries}a; magnitudes shown in the correct relative order but offset for clarity). The \textit{Gaia} light curve shows small differences in the depths of the minima, typical of a system in which the two stars have different masses \citep{boffinjones2019book}.

\textit{JaSt 2--4:} this is a faint, slightly elliptical ring PN (PN G001.0+01.4) discovered by \citet{jacoby2004jast24}. The \textit{Gaia} photometric data of its CSPN
(\textit{Gaia} DR3 4060890513224820352, $\alpha = 17^h42^m28.06^s\ \delta = -27^\circ13^{\prime\prime}31.79^\prime$)
shows ellipsoidal modulation effects corresponding to an orbital period of $\sim$11.2 hours (Fig.~\ref{fig:binaries}b; magnitudes are offset).

\textit{IPHASX J192534.9+200334:} this is an irregularly shaped ``likely" PN (PN G054.5+01.8) discovered by \citet{sabin2014iphaspne}. Its CSPN is also missing from \citet{chornay2021cspn}, but the \textit{Gaia} DR3 data for 
the closest source to the center of the nebula
(\textit{Gaia} DR3 4515887189511585792, $\alpha = 19^h25^m34.90^s\ \delta = +20^\circ03^{\prime\prime}34.74^\prime$)
shows that it is indeed the CSPN, a doubly eclipsing binary system with a $\sim$10.2 hour period and sinusoidal brightness variability due to irradiation of the cooler companion (Fig.~\ref{fig:binaries}c).

\textit{Pa 164:} this is a ``likely" PN (PN G061.5--02.6) from \citet{kronberger2017}. The \textit{Gaia} light curve of its CSPN
(\textit{Gaia} DR3 1834171384397003264, $\alpha = 19^h57^m23.23^s\ \delta = +23^\circ52^{\prime\prime}48.27^\prime$)
shows it to be an eclipsing irradiated binary with a period of $\sim$28 hours (Fig.~\ref{fig:binaries}d).
A secondary eclipse is not evident in the \textit{Gaia} data. It was published as a candidate variable by \citep{chornay2021gaiabinaries}.

\section{Conclusion}

We have found four new close binary CSPNe using the \textit{Gaia} DR3 epoch photometry data.
Two of the sources had not been previously identified as CSPNe, because of the extended natures of their nebulae and their lack of obvious blue colors expected for CSPNe. This highlights a difficulty in CSPN identification especially applicable to close binaries, where a significant contribution to the flux can come from the cooler companion star. 

We note that about a quarter of the previously known close binary CSPN population (26 objects) have matching sources classified as variable in \textit{Gaia} DR3. Our analysis recovers the correct period (or half of it, in the case of ellipsoidal modulation) for 21 of these objects. For three of these objects the shape of folded light curve is not obviously due to binarity, despite the period being correct. Thus it is very likely that there are more close binaries waiting to be discovered in the \textit{Gaia} photometry, though these will benefit from ground-based confirmation.

\textit{Gaia} DR3 is based on 34 months of data. The duration of the data collection will be nearly doubled in the next data release, with the full mission duration expected to be over ten years.\footnote{\url{https://www.cosmos.esa.int/web/gaia/release}} Future releases will thus benefit from a longer baseline and many more data points \citep{eyer2022variablesarxiv}, which will continue to make \textit{Gaia} a valuable tool for probing CSPN variability. This is particularly true for systems where bright nebulae or nearby stars complicate ground-based photometric monitoring.

\begin{acknowledgments}
This research has made use of data from the European Space Agency (ESA) mission \textit{Gaia} (\url{https://www.cosmos.esa.int/gaia}) processed by the {\it Gaia} Data Processing and Analysis Consortium (DPAC, \url{https://www.cosmos.esa.int/web/gaia/dpac/consortium}). Funding for the DPAC has been provided by national institutions, in particular the institutions participating in the {\it Gaia} Multilateral Agreement.

This research was supported through the Cancer Research UK grant A24042.
\end{acknowledgments}

%

\vspace{5mm}
\facilities{Gaia}


\software{astropy \citep{2013A&A...558A..33A,2018AJ....156..123A}. 
}

\bibliography{bib}{}
\bibliographystyle{aasjournal}

\end{document}